\documentstyle{elsart}

\input epsf.tex

\newcommand{\f}{\frac}

\begin{document}
\begin{frontmatter}
\title{Gauge invariance of the $\beta$-function \\
       in nonrelativistic quantum electrodynamics}
\author{Olav F. Sylju{\aa}sen\thanksref{email}}
\thanks[email]{E-mail: sylju@physics.ucla.edu. Fax: 310-825-5734.}
\address{Department of Physics and Astronomy,
         University of California Los Angeles, \\
         Los Angeles, California 90095-1547, USA}

\begin{abstract}
\baselineskip=16pt
Recently there has been much interest in gauge theories applied to
condensed matter physics. I show that for a system of 
nonrelativistic electrons coupled to a U(1) gauge field in
the presence of a Fermi surface, the $\beta$-function to one-loop order is,
for a particular family of gauge-choices, independent of the gauge-choice. 
\end{abstract}   
\end{frontmatter}
\section{Introduction}
That gauge fields successfully describe the forces between elementary
particles is certainly among the greatest discoveries in physics.
But gauge theories are not restricted
to high energy physics. 
In condensed matter physics one is often interested in theories
with constraints. In a sense
a constraint is just a (infinitely strong) force and
it is quite natural that these theories also can be formulated
as gauge theories.
\\
\\
The gauge field description is in terms of
gauge {\em potentials}, which are redundant variables.   
The theory is therefore invariant with respect to
a change of gauge. The most important consequence
of this gauge symmetry is that it dictates the form of the
interaction vertices between the gauge fields and the other
fields in the theory. Other than that; the gauge-invariance is just
a reminder that we are working with redundant variables. This
redundancy can be very annoying, in particular when using 
path-integrals, and it is often necessary to choose a specific
gauge. This choice of gauge {\em cannot} affect any physical quantities
calculated in the theory.    
\\
\\
In this paper we will consider a theory of $U(1)$ gauge bosons
coupled to nonrelativistic fermions. This theory is motivated by
the composite fermion approach to the half-filled
Landau level state in the Quantum Hall Effect.
\\
\\
Some years ago it was observed that the $\nu=1/2$
Quantum Hall state exhibits metal-like features \cite{HWJ}.
Using the idea of Jain \cite{Jain}, in which two flux quanta
are "attached" to each electron; making a composite fermion, 
Halperin,Lee and Read \cite{HLR} formulated the theory of the
$\nu=1/2$ state as a theory of composite fermions interacting with
a fictitious gauge field, in {\em zero}
external magnetic field. In the mean field approximation, in which
the electrons are uniformly spread out, this theory is just that
of free (composite) fermions. Therefore, in the mean field
approximation, 
the $\nu=1/2$ state is certainly metallic. 
Striking evidence of the existence of 
a Fermi surface was subsequently observed experimentally
\cite{SAW},\cite{Antidot},\cite{Focusing}.\\
\\
In order to make a consistent theory of the $\nu=1/2$ state it is
necessary to also include the effects of gauge-fluctuations.
However, when fluctuations of the gauge field 
are taken into account, it seems
at first sight that the mean field picture is destroyed.
As shown in \cite{HLR} the self-energy correction due to the
transverse gauge fluctuations dominates over the linear 
in frequency Fermi liquid term at low frequencies. 
\\
\\
In an attempt to address the role of the gauge field fluctuations
Nayak and Wilczek \cite{NW} showed, using a renormalization-group approach,
that for electron-electron interactions of shorter range than $1/r$, 
the fluctuations lead to a strong coupling fixed point in the infra-red, different from the Fermi liquid fixed point.
Their result was based on the calculation of the single-particle
Green function and gauge-invariance through the Ward identity.
A somewhat similar approach was taken by Chakravarty {\em et al} \cite{CNS}.
They studied the same system using the $\epsilon$-expansion around $d=3$,
and obtained essentially the same result; a strong coupling fixed point.
\\
\\
Stern and Halperin \cite{Stern} constructed a Boltzman transport equation
for the $\nu=1/2$ state
in the same way as Landau did for the Fermi liquid.
This construction is based on the pole structure of the
single-particle Green function. 
\\
\\
As emphasized in \cite{KFWL} the single-particle
Green function is not gauge-invariant, and so any
conclusions based on it might not be physical.  
This leads naturally to the question
whether or not the existence of the strong coupling fixed point found
in \cite{NW},\cite{CNS}, or the pole structure used 
in \cite{Stern} is dependent on the choice of gauge?
\\
\\
We will in this paper show that although the single-particle Green function
is dependent on the choice of gauge, the $\beta$-function is not, i.e.
neither the existence of the strong 
coupling fixed point found by Nayak and Wilczek
nor the pole structure used by Stern and Halperin is
dependent on the choice of gauge. 

\section{Calculations in d=2}
The Lagrangian density considered in this paper is
\begin{eqnarray}
   {\cal L} & = & \psi^{\dagger} \left( \partial_0 - \mu \right) \psi
   -\f{1}{2m} \psi^{\dagger} \left( \partial_i - iga_i \right)^2 \psi
    \nonumber \\
    &   &
   +\f{1}{2} \int d^dy \nabla \times \vec{a} \: V(|\vec{x}-\vec{y}|) 
    \nabla \times \vec{a}
   +\f{1}{2\alpha} \left( \nabla \cdot \vec{a} \right)^2. \label{lagrangian}
\end{eqnarray}
This Lagrangian is motivated by the composite fermion description
of the $\nu=1/2$ Quantum Hall state, where two flux
tubes are attached to each electron forming composite fermions.
Since the flux quanta are attached to
the electrons, fluctuations in electron-density lead to fluctuations
in the gauge field. This is taken care of by the third term in which the
electron-electron interaction $V(|\vec{x}-\vec{y}|)$ controls the gauge fluctuations.
Here we will consider electron-electron interactions of the form
\begin{equation}
   V(|\vec{x}-\vec{y}|) \propto \f{1}{|\vec{x}-\vec{y}|^{\eta}},
\end{equation}
where $\eta$ is close to $1$.
The last term is the gauge-fixing term. It is obtained using the
Faddeev-Popov Ansatz to restrict $\vec{a}$ to a gauge
in which
\begin{equation}
     \nabla \cdot \vec{a} = f,
\end{equation}
and then averaging over a Gaussian distribution, with mean value $0$ and
width $\alpha$, of such functions $f$. 
Coulomb gauge corresponds to $\alpha \rightarrow 0$. \\
\\
In order to describe the $\nu=1/2$ state properly, a Chern-Simons
term and a time-like gauge field ($a_0$) should also be included.
We omit these terms because it is likely, as for Coulomb 
interaction in metals, that the $a_0$-field is screened and does
not contribute to the fluctuations at low frequencies. 
There are however subtleties associated with the fluctuations
of the $a_0$ field \cite{HLR},\cite{LKGK} which we do not address here.
\\ 
\\
Before we go on to study the Lagrangian (\ref{lagrangian}), let us 
understand why
the $\beta$-function in QED is gauge-invariant.
The QED-Lagrangian density in terms of the bare coupling and fields is
\begin{equation} \label{QED}
 {\cal L} =-\f{1}{4} F_0^{\mu \nu} F^0_{\mu \nu}
           +\bar{\psi}_0 \left( \FMSlash{\partial} 
                        -ie_0 \FMSlash{A}_0 -m_0 \right) \psi_0.
\end{equation}
By redefining fields and the coupling constant
\begin{equation} \label{QEDren}
 {\cal L} =-\f{1}{4} Z_3 F^{\mu \nu} F_{\mu \nu}
           + Z \bar{\psi} \left(\FMSlash{\partial} 
                        -\sqrt{\f{Z_3}{Z_e}} ie \FMSlash{A}-Z_m m 
                        \right) \psi.
\end{equation}
The form of the interaction between the fermions and the gauge-field
is dictated by gauge-invariance. So for gauge-invariance to hold in
the renormalized theory:
\begin{equation} \label{Ward}
      Z_3 = Z_e. 
\end{equation}
This identity can also be proven using the Ward identity, which is nothing 
but the statement of gauge-invariance. 
Now consider the correlation function
\begin{equation} \label{corr}
   \left< F^{\mu \nu}(x) F_{\alpha \beta}(y) \right>.
\end{equation}
This correlation function is obviously gauge-independent, and therefore
$Z_3$ should also be independent of gauge. This and the above Ward
identity (\ref{Ward}) ensures that the $\beta$-function which is 
obtained by differentiating $Z_e$ is independent of the choice
of gauge. 
\\
\\
We see that neither $Z$ nor $Z_m$ play any role in this
argument. Let us think about computing the above correlation function 
(\ref{corr}).
The exact answer will of course be a very complicated function
of $m_0=Z_m m$ and $e_0=e Z_e^{-1/2}$. This answer should not 
change as we change the gauge, and so if $Z_m$ and $Z_e$ were
dependent on gauge, amazing cancellations would have to occur.
Conceivably the only cancellations that can occur should
be due to the Ward identity. Since the Ward identity has nothing
to do with $Z_m$ it follows that $Z_m$ cannot be dependent on gauge
either. $Z$ is not constrained by the above as
the correlation function (\ref{corr}) can be 
calculated using the effective action for the gauge fields in
which the fermions are integrated out. $Z$ will be absorbed 
into the measure,
and there is nothing preventing $Z$ from being gauge-dependent.
\\
\\
For non-abelian gauge theories the situation is somewhat different from 
QED. There the gauge-field renormalization constant is dependent on gauge,
but in spite of that, the $\beta$-function is gauge-independent there also.
\\
\\
Let us now return to the Lagrangian (\ref{lagrangian}). 
Written in Fourier space
the Euclidean action in terms of renormalized couplings and fields is
\begin{eqnarray}
S & = & T \sum_n \int \f{d^d p}{(2\pi)^d} 
          Z \psi^{\dagger} (p) 
          \left( -\mathrm{i} \omega_n +Z_m \left( \f{p^2}{2m}-\mu \right)
          \right) \psi (p) 
          \nonumber \\
  &   &+T \sum_m \int \f{d^d q}{(2\pi)^d}
          Z_3 \f{1}{2} a^i(q) 
          \left( \vec{q \:}^2 V(q) \left( \delta^{ij} -\f{q^i q^j}{q^2} \right)
                 +Z_{\alpha} \f{1}{\alpha} q^i q^j \right)
           a^j(-q)
           \label{lagren} \\
  &   &-Z Z_m\sqrt{\f{Z_3}{Z_g}} \f{g}{2m}  T \sum_n \int \f{d^d p}{(2\pi)^d} 
                   T \sum_m \int \f{d^d q}{(2\pi)^d}
        \left( 2p^i+q^i \right) a_i(q)
        \psi^{\dagger}(q+p)
        \psi (p) \nonumber \\
  &   &+Z Z_m \f{Z_3}{Z_g} 
       \f{g^2}{2m} T \sum_n \int \f{d^d p}{(2\pi)^d} 
       T \sum_{m_1} \int \f{d^d q_1}{(2\pi)^d}
       T \sum_{m_2} \int \f{d^d q_2}{(2\pi)^d}
       \nonumber \\
  &   &\times \psi^{\dagger} (p+q_1+q_2)
       \psi (p) a^i(q_1) a^i(q_2). \nonumber
\end{eqnarray}
We will assume a circular Fermi surface and eventually take
the zero- temperature limit such that $\mu=p_f^2/2m$.
It is of course possible to linearize the energy dispersion around
the Fermi surface, in that case $Z_m$ is naturally named $Z_{v_f}$.
This linearization cannot be essential for any important
physics and will not be used in this paper. Note that there is no renormalization constant associated
with $p_f$. $p_f$ measures the total number of particles
which does not change. 
The renormalization of $V(q)=v_B q^{\eta-2}$
is taken care of by $Z_3$.
\\
\\
By simple dimensional analysis one can verify that  
the relevant coupling between the gauge field and the fermions
is $g^2 v_f p_f^{d-1-\eta}/v_B$ and not $g^2$ as in QED. 
Therefore the relevant $\beta$-function for this problem doesn't involve
just $Z_g$, but $Z_g$,$Z_m$ and $Z_3$.
As in QED, gauge invariance requires $Z_3=Z_g$. 
This can be seen explicitly by writing down the Ward identity. Our
Lagrangian is not gauge-invariant with respect to gauge-
transformations which mix space and time. We must therefore consider
the spatial vertex separately in the Ward identity, 
setting the external frequency to zero.
As $Z_3$ is obviously gauge-invariant, we need only to find 
if $Z_m$ depends on the choice of gauge, to make
statements about the $\beta$-function.
\\
\\
In the mean field approximation the gauge field vanishes. 
Since the mean field theory describes a metal it is reasonable to 
proceed as in the theory of metals where one uses the RPA-corrected form
of the interaction. 
The bare photon propagator extracted from the Lagrangian is
\begin{equation} 
  D_0^{ij} (q,\epsilon_m) = \f{1}{V(q)q^2} 
                            \left( \delta^{ij} -\f{q^i q^j}{q^2} \right)
                            +\alpha \f{q^i q^j}{(q^2)^2}.
\end{equation}
The RPA-approximation of the propagator is obtained by summing
an infinite series of bubble diagrams. 
\begin{figure}
\begin{center}
\leavevmode
\epsfbox{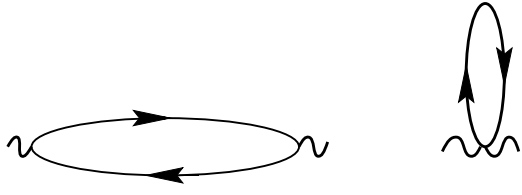}
\caption{One-loop diagrams used in the RPA resummation.}
\end{center}
\end{figure}
The polarization bubble is
shown in figure 1, and its expression is
\begin{equation}
  \Pi^{ij} (q,\epsilon_m) = A(q,\epsilon_m) 
                            \left(\delta^{ij} -\f{q^i q^j}{q^2} \right)
                           +B(q,\epsilon_m) \f{q^i q^j}{q^2}
\end{equation}
where
\begin{eqnarray}
   A(q,\epsilon_m) & = & -\left( \f{g}{2m} \right)^2 \f{4}{d-1}
                         T \sum_n \int \f{d^d p}{(2\pi)^d}
                         (\vec{p} \times \vec{q})^2 G_0(\vec{p},\omega_n)
                         G_0(\vec{p}+\vec{q},\omega_n+\epsilon_m)
                         \nonumber \\
                   &   & +\f{g^2}{m} T \sum_n \int \f{d^d p}{(2\pi)^d}
                         G_0(\vec{p},\omega_n), \\
   B(q,\epsilon_m) & = & -\left( \f{g}{2m} \right)^2 
                         T \sum_n \int \f{d^d p}{(2\pi)^d}
                         \f{(2\vec{p} \cdot \vec{q}+\vec{q}^2)^2}{\vec{q}^2}
                         G_0(\vec{p},\omega_n)
                         G_0(\vec{p}+\vec{q},\omega_n+\epsilon_m)
                         \nonumber \\
                   &   & +\f{g^2}{m} T \sum_n \int \f{d^d p}{(2\pi)^d}
                         G_0(\vec{p},\omega_n),
\end{eqnarray}
and the bare single-particle propagator is
\begin{equation}
   G_0(\vec{p},\omega_n) = \f{-1}{i \omega_n -   
   (\f{\vec{p}^2}{2m}-\mu)}.
\end{equation}
Summing the infinite series of bubble diagrams we obtain
\begin{equation} \label{RPAgauge}
   D^{ij} (q,\epsilon_m) = \f{1}{V(q)q^2-A(q,\epsilon_m)}
                           \left( \delta^{ij} - \f{q^i q^j}{q^2} \right)
                          +\f{\alpha}{q^2-\alpha B(q,\epsilon_m)}
                           \f{q^i q^j}{q^2}.
\end{equation}
The transverse part is independent of the gauge-fixing
parameter $\alpha$, and the longitudinal part of the gauge interaction
does not include any information about the electron-electron interaction.
\\
\\
The calculations of $A(q,\epsilon_m)$ and $B(q,\epsilon_m)$ are 
straight-forward at $T=0$. In the zero-temperature limit of the imaginary time
formalism we get in $d=2$, for $|\epsilon| \ll v_f q$ and $q \ll p_f$:
\begin{eqnarray}
   A(q,\epsilon) & = & -\f{g^2}{12 \pi} \f{q^2}{2m} 
                       -\f{g^2}{\pi} \f{p_f^2}{2m} \f{|\epsilon|}{v_f q}
                       ,
                       \label{d2A}     \\
   B(q,\epsilon) & = & -\f{g^2}{\pi} \f{p_f^2}{2m} 
                        \left( \f{\epsilon}{v_f q} \right)^2. 
                       \label{d2B} 
\end{eqnarray}
Having calculated the gauge boson propagator we can now calculate 
the self-energy diagrams.
\begin{figure}
\begin{center}
\leavevmode
\epsfbox{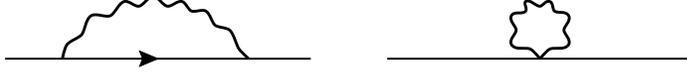}
\caption{One-loop diagrams for the self-energy.}
\end{center}
\end{figure}

The one-loop diagrams contributing to the self-energy are drawn in
figure 2. The expression is
\begin{eqnarray}
   \Sigma(\vec{p},\omega_n) & = & \left( \f{g}{2m} \right)^2
                        T\sum_m \int \f{d^dq}{(2\pi)^d} \left( 2p^i+q^i
                        \right) D^{ij} (q,\epsilon_m)   
                        \left( 2p^j+q^j \right) G_0(\vec{p}+\vec{q},\omega_n
                        +\epsilon_m) 
                        \nonumber \\
                        &   & -\f{g^2}{2m} T \sum_m
                        \int \f{d^dq}{(2\pi)^d} D^{ii} (q,\epsilon_m).
\end{eqnarray}
For convenience we will split the contributions to the self-energy into two pieces,
\begin{equation}
 \Sigma(\vec{p},\omega_n)  =  \Sigma_t(\vec{p},\omega_n)
                               +\Sigma_l(\vec{p},\omega_n).
\end{equation}
The first part is the self-energy obtained from the diagrams in
figure 2 using the transverse part of the gauge propagator, while
the second part is obtained using the longitudinal propagator in the
same diagrams. It is obvious from the form of the gauge-propagator 
(\ref{RPAgauge}) that
only $\Sigma_l$ will depend on gauge. Let us first show that the 
logarithmic singularity necessary for the strong coupling fixed point 
is obtained by calculating $\Sigma_t$.
Performing the angular integral in $d=2$,
the transverse contribution to the one-loop self-energy can be written
\begin{eqnarray}
    \Sigma_t(\vec{p},\omega_n) & = &-\f{g^2}{2 \pi}
                           T \sum_m \int_0^\Lambda dq
                           \f{q^{-1}}{V(q)q^2-A(q,\epsilon_m)}
                           \left[ \mathrm{i} \omega_n - \f{p^2}{2m}+\mu 
                           \right. \\
                      &   & \left. 
                         - \mathrm{i} \mathrm{sgn} \left( \omega_n+\epsilon_m \right)
                           \left( \f{p^2q^2}{m^2}
                               -\left( \mathrm{i} \omega_n-\f{p^2}{2m}+\mu+
                                       \mathrm{i} \epsilon_m -\f{q^2}{2m} 
                           \right)^2 \right)^{1/2} \right]. \nonumber
\end{eqnarray}
We have inserted an arbitrary cutoff $\Lambda$ in the
$q$-integration and
used the fact that $A(q,-\mathrm{i} \epsilon_m)=A(q,\mathrm{i} \epsilon_m)$.
Expanding the square root the self-energy takes the form
\begin{eqnarray}
  \Sigma_t(\vec{p},\omega_n) & = & -\f{g^2}{2 \pi}
                           T \sum_m \int_0^\Lambda \f{dq}{q}
                           \f{1}{V(q)q^2-A(q,\epsilon_m)} \nonumber \\
       &  & \times \left[ \left( \mathrm{i} \omega_n - \f{p^2}{2m}+\mu \right)
                  \left( 1+ i\mathrm{sgn}(\omega_n+\epsilon_m) \f{m}{pq} 
                            ( \mathrm{i} \epsilon_m-\f{q^2}{2m} ) +
                            \cdots \right) \right. \nonumber \\
       &  & -i \mathrm{sgn}( \omega_n+\epsilon_m) \f{pq}{m} 
                     \left( 1- \f{1}{2} ( \f{m}{pq} )^2 (
                     \mathrm{i} \epsilon_m -\f{q^2}{2m} )^2+\cdots \right)
                     \nonumber \\
       &  & \left. +O \left(\mathrm{i} \omega_n-\f{p^2}{2m}+\mu \right)^2 \right].
            \label{above}
\end{eqnarray}
The ellipses indicate higher order terms in the expansion of the square root. 
These terms are not as singular as the leading term, so they can
safely be neglected.
Let us now expand the renormalization constants $Z$ and $Z_m$
in the coupling constant,
\begin{eqnarray}
     Z & = & 1 + Z^{(1)} + \cdots, \\
     Z_m & = & 1+ Z^{(1)}_m + \cdots.
\end{eqnarray}
Using this expansion we have
\begin{eqnarray}
    Z \left( \mathrm{i} \omega_n-Z_m ( \f{p^2}{2m} -\mu ) \right)
     & = & \left(1+Z^{(1)}+Z^{(1)}_m \right) 
      \left( \mathrm{i} \omega_n- ( \f{p^2}{2m} -\mu ) \right) \nonumber
      \\
     &   &  -Z^{(1)}_m \mathrm{i} \omega_n+\cdots.
\end{eqnarray}
Contributions to $Z^{(1)}_m$ are 
distinguished by the fact that they renormalize $\mathrm{i} \omega_n$ differently
from $(p^2/2m-\mu)$. So it is the second term
in the above expression (\ref{above}) which will contribute to $Z^{(1)}_m$.
The second term is
\begin{eqnarray}
   \Sigma_t^{Z_m} (\vec{p},\omega_n)
                     & = & \mathrm{i} \f{g^2}{2\pi} \f{p}{m} T \sum_m 
                   \int_0^\Lambda dq \f{\mathrm{sgn}
                    (\omega_n+\epsilon_m)}{V(q)q^2-A(q,\epsilon_m)}
                   \nonumber \\
                   &   &\times
                   \left( 1-\f{1}{2} ( \f{m}{pq} )^2
                   ( \mathrm{i} \epsilon_m-\f{q^2}{2m} )^2 
                   +\cdots \right). 
\end{eqnarray}
When $\omega \rightarrow 0$ the first term in the last
parenthesis will give the leading behaviour. Setting $T=0$,
$p=p_f$, $V(q)=v_B q^{\eta-2}$ and 
using the result (\ref{d2A}) for $A(q,\epsilon)$ this leading behaviour is
\begin{eqnarray}
    \Sigma_t^{Z_m} (p_f,\omega) & = & 
                  \mathrm{i} \f{g^2}{2\pi} v_f \int_{-\infty}^{\infty} 
                    \f{d\epsilon}{2\pi}
                 \int_0^\Lambda dq 
                 \f{\mathrm{sgn}(\omega+\epsilon)}{v_B q^{\eta}+
                       \f{g^2}{12 \pi} \f{q^2}{2m} 
                       +\f{g^2}{\pi} \f{p_f^2}{2m} \f{|\epsilon|}{v_f q}}
                 \nonumber \\
           & \approx & -\mathrm{i}
            \omega\f{g^2 v_f \Lambda^{1-\eta}}{2\pi^2 v_B} 
                         \f{\left(\f{g^2 v_f \Lambda^{1-\eta}}{2\pi^2 v_B}
                                \f{\pi |\omega|}{p_f v_f} 
                                \left( \f{p_f}{\Lambda} \right)^2
                                \right)^{\f{1-\eta}{1+\eta}}}{1-\eta}.
\end{eqnarray}
In the special case $\eta=1$,
corresponding to $1/r$-interactions between the fermions, we find
in the low-frequency limit: 
\begin{eqnarray}
  \Sigma_t^{Z_m} (p_f,\omega) & = & -\mathrm{i} \omega \f{g^2 v_f}{4\pi^2 v_B} 
                   \ln \left( \f{g^2 v_f}{2\pi^2 v_B} 
                               \f{\pi |\omega|}{v_f p_f}
                               \left( \f{p_f}{\Lambda} \right)^2 \right).  \end{eqnarray}
In this case the correction to 
the $\mathrm{i} \omega$-behaviour is logarithmic, as observed by Nayak and
Wilczek. This case is the marginal case which sets the borderline
between the
Fermi liquid theory and a strong coupling theory.
For $\eta < 1$, corresponding to interactions more long-range
than Coulomb interactions, the self-energy vanishes faster than $\omega$ and
no deviations from Fermi liquid are seen. But for $\eta > 1$, short-range
electron-electron interactions, the
self-energy becomes the dominant term at small $\omega$ and the 
system is not in the Fermi liquid universality class.\\                              \\
The question is now whether these conclusions depend on the choice
of gauge. At first sight they do not, since the transverse gauge-propagator
does not depend on $\alpha$. However, one could imagine that the longitudinal
gauge-propagator, which is dependent on $\alpha$, could give additional
competing singularities. We will now show, by computing
the contribution to the self-energy using the longitudinal 
gauge-propagator, that there are no $\alpha$-dependent competing
singularities.\\
\\ 
Performing the angular integration in the expression
for the self-energy contribution coming from the longitudinal gauge 
propagator we get                    
\begin{eqnarray}
    \Sigma_l(\vec{p},\omega_n) & = & +\f{g^2}{2\pi}
                            T \sum_m \int_0^\Lambda dq
                           \f{\alpha q^{-1}}{q^2-\alpha B(q,\epsilon_m)}
                           \left[ \mathrm{i} \omega_n - \f{p^2}{2m}+\mu 
                           \right. \nonumber \\
                        &   & \left.
                           +\mathrm{i}
                            \f{\mathrm{sgn} \left( \omega_n+\epsilon_m \right)
                             \left( \mathrm{i} \omega_n -\f{p^2}{2m}+\mu +\mathrm{i} \epsilon_m
                            \right)^2}
                            {\sqrt{\left( \f{pq}{m} \right)^2 - 
                                   \left( \mathrm{i} \omega_n -\f{p^2}{2m} +\mu 
                                          +\mathrm{i} \epsilon_m -\f{q^2}{2m}
                                   \right)^2}} \right].
\end{eqnarray}
The gauge boson propagator is invariant with respect to $\epsilon_m
\rightarrow -\epsilon_m$. Therefore no gap in the fermion spectrum 
is generated by the self-energy,
which of course is expected on general grounds. In the same way as
for the transverse contribution we expand the square root.
\begin{eqnarray}
    \Sigma_l(\vec{p},\omega_n) & = & \f{g^2}{2\pi}
                            T \sum_m \int_0^\Lambda dq
                           \f{\alpha q^{-1}}{q^2-\alpha B(q,\epsilon_m)}
                           \left[ \left( \mathrm{i} \omega_n - \f{p^2}{2m}+\mu \right)
                           \right. \nonumber \\
                  &   &    \times \left(1+
                           2  \mathrm{i}\mathrm{sgn}(\omega_n+\epsilon_m) 
                                        \f{m}{pq} \mathrm{i}\epsilon_m+\cdots
                                  \right) \nonumber \\
                   &   &  +\mathrm{i} \mathrm{sgn}
                           ( \omega_n+\epsilon_m ) (\mathrm{i} \epsilon_m)^2
                             \f{m}{pq} \left( 1+\f{1}{2} 
                                       \left( \f{m}{pq} \right)^2
                                       \left( \mathrm{i} \epsilon_m-
                                       \f{q^2}{2m} \right)^2
                                       +\cdots \right)
                                  \nonumber \\
                   &    & \left. +O \left( \mathrm{i} \omega_n-\f{p^2}{2m}+\mu \right)^2
                          \right]. 
\end{eqnarray}                          
The contribution to $Z_m$ comes from the second term and is
\begin{eqnarray}
   \Sigma_l^{Z_m} (\vec{p},\omega_n) & = & \f{g^2}{2\pi}
                            T \sum_m \int_0^\Lambda dq
                           \f{\alpha q^{-1} 
                  \mathrm{i} \mathrm{sgn}( \omega_n+\epsilon_m ) }
                             {q^2-\alpha B(q,\epsilon_m)}
                        (\mathrm{i} \epsilon_m)^2
                             \f{m}{pq} \left( 1+\cdots \right).
\end{eqnarray}
This quantity does not vary fast with $p$ so we set $p=p_f$. Taking
the zero-temperature limit,using the result (\ref{d2B}) for $B(q,\epsilon)$
 and keeping just the dominant term as $\omega \rightarrow 0$ we get
\begin{eqnarray}
   \Sigma_l^{Z_m} (p_f,\omega) & = & \f{g^2}{4\pi^2}
                            \int_{-\infty}^{\infty} d\epsilon
                           \int_0^\Lambda dq
                           \f{\alpha q^{-1}}{q^2+
                              \alpha \f{g^2}{\pi} \f{p_f^2}{2m}
                                  \f{\epsilon^2}{v_f^2 q^2}}
                          \mathrm{i} \mathrm{sgn}( \omega+\epsilon) 
                         (\mathrm{i} \epsilon)^2 \f{m}{pq} 
                          \nonumber \\
                         & \approx & -\mathrm{i} \omega 
                                  \f{\sqrt{2}}{6}
                                  \left( \f{\alpha g^2 v_f}{2 \pi p_f} \right)^{1/4}
                                  \left( \f{|\omega|}{v_f p_f} \right)^{1/2}
                                 \nonumber \\
                         & \sim & \omega^{3/2},
\end{eqnarray}
which vanishes faster than $\omega$ and therefore does not contribute to
the singularities of $Z_m$. 
This shows that the singularities of $Z_m$ are not dependent on $\alpha$
to this order in perturbation theory. 
\newpage
\section{Calculations in d=3}
In \cite{CNS}  the authors used the $\epsilon$-expansion around $d=3$,
and the free gauge-propagator 
\begin{equation} \label{freegauge}
   D^{ij} = \f{1}{v_B q^2} \left( \delta^{ij} -\f{q^i q^j}{q^2} \right),
\end{equation}
to analyze the same theory.
We will in this chapter show that the $\beta$-function
is independent of gauge also in this case. 
The computation in three dimensions follows along the same lines as
the computations in $d=2$. 
The polarization diagrams at $T=0$ give for $|\epsilon| \ll v_f q$ and
$q \ll p_f$,
\begin{eqnarray}
   A(q,\epsilon) & = & -\f{g^2}{4 \pi^2} p_f \f{p_f^2}{2m} \left( \f{q}{p_f}
                        \right)^2 
                       -\f{g^2}{4 \pi} p_f \f{p_f^2}{2m} \f{|\epsilon|}{v_f q}
                       ,
                       \label{d3A} \\
   B(q,\epsilon) & = &  -\f{g^2}{\pi^2} p_f \f{p_f^2}{2m} 
                        \left( \f{\epsilon}{v_f q} \right)^2.
                       \label{d3B} 
\end{eqnarray}
The contribution
from the transverse gauge boson to the self-energy takes the following
form when the angular integrations are performed
\begin{eqnarray}
 \Sigma_t(\vec{p},\omega_n)  
           & = & 
          -\f{g^2}{4 \pi^2} 
           T \sum_m \int_0^\Lambda dq \f{1}{V(q)q^2-A(q,\epsilon_m)}
           \left[ 
           2\left( \mathrm{i} \omega_n-\f{p^2}{2m}+\mu +
                   \mathrm{i} \epsilon_m +\f{q^2}{2m} \right)
            \right. \nonumber \\
             &   & +\f{m}{pq} \left( 
         \left( \mathrm{i} \omega_n-\f{p^2}{2m}+\mu+
         \mathrm{i} \epsilon_m-\f{q^2}{2m} \right)^2
         - \f{p^2q^2}{m^2} \right) \nonumber \\
       &   & \times \left.
         \ln \left[ \f{\f{pq}{m}- 
             \left( \mathrm{i} \omega_n-\f{p^2}{2m}+\mu+
             \mathrm{i} \epsilon_m-\f{q^2}{2m} \right)
             }{ -\f{pq}{m}- 
         \left( \mathrm{i} \omega_n-\f{p^2}{2m}+\mu+
         \mathrm{i} \epsilon_m-\f{q^2}{2m} \right)}
         \right] \right]. 
\end{eqnarray}
For small $\omega_n$, $p^2/2m-\mu$ and $\epsilon_m \ll v_f q$,
$q \ll p_f$  the logarithm can be expanded;
\begin{eqnarray}
   \ln \left[ \f{\f{pq}{m}- 
             \left( \mathrm{i} \omega_n-\f{p^2}{2m}+\mu+\mathrm{i} \epsilon_m-\f{q^2}{2m} \right)
             }{ -\f{pq}{m}- 
         \left( \mathrm{i} \omega_n-\f{p^2}{2m}+\mu+\mathrm{i} \epsilon_m-\f{q^2}{2m} \right)}
         \right]
    & = &\mathrm{i} \pi \mathrm{sgn}(\omega_n+\epsilon_m) \\
    &   &-\f{2m}{pq} 
       \left( \mathrm{i} \omega_n-\f{p^2}{2m}+\mu+\mathrm{i} \epsilon_m-\f{q^2}{2m} \right)+\cdots.
       \nonumber 
\end{eqnarray}
When using this expansion in $\Sigma_t$ we can write
\begin{eqnarray}
  \Sigma_t(\vec{p},\omega_n) & = & -\f{g^2}{4 \pi^2} 
           T \sum_m \int_0^\Lambda dq \f{1}{V(q)q^2-A(q,\epsilon_m)}
  \left[ \left( \mathrm{i} \omega_n-\f{p^2}{2m}+\mu \right) \right. \nonumber \\
         &   & \times
           \left( 4+2\f{m}{pq} (\mathrm{i} \epsilon_m-\f{q^2}{2m} ) 
                \mathrm{i} \pi
                \mathrm{sgn}(\omega_n+\epsilon_m) +\cdots \right)
                \nonumber \\
             &   & + \f{pq}{m} \mathrm{i} \pi \mathrm{sgn}(\omega_n+\epsilon_m)
                  +4\mathrm{i} \epsilon_m +\cdots \nonumber \\
             &   &\left. +O\left( \mathrm{i} \omega_n-\f{p^2}{2m}+\mu \right)^2
                  \right].
\end{eqnarray}
Using the expression (\ref{d3A}) for $A(q,\epsilon_m)$ all terms
odd in $\epsilon_m$ will vanish, and so no mass term is generated.
The most singular term contributing to $Z_m$ is
\begin{equation}
    \Sigma^{Z_m}_t(\vec{p},\omega_n) =
                 \f{g^2}{4\pi^2} \f{p}{m}
                  T \sum_m \int_0^\Lambda dq \f{q i\pi
                     \mathrm{sgn}(\omega_n+\epsilon_m)}{
                     q^2 V(q)-A(q,\epsilon_m)},
\end{equation}
Taking the zero-temperature limit and setting $p=p_f$ we get for 
the dominant term in $\omega$ as $\omega \rightarrow 0$:
\begin{equation}
    \Sigma^{Z_m}_t(p_f,\omega_n) =
                 \f{g^2 v_f}{4\pi^2} \mathrm{i} \omega
                 \int_0^\Lambda dq \f{q}{
                     q^2 V(q)+\f{g^2}{4 \pi^2} p_f \f{p_f^2}{2m} 
                     \left( \f{q}{p_f} \right)^2 +
                 \f{g^2}{4 \pi} p_f \f{p_f^2}{2m} \f{|\omega|}{v_f q}},
\end{equation}
where we have used the expression (\ref{d3A}) for $A(q,\epsilon)$.
For the free gauge-propagator (\ref{freegauge}), $V(q)=v_B$, and 
the leading behaviour as $\omega \rightarrow 0$ is
\begin{equation}
    \Sigma^{Z_m}_t(p_f,\omega_n) =
                 -\mathrm{i} \omega
                  \f{g^2 v_f}{12\pi^2 v_B}
                 \ln \left( \f{g^2 v_f}{8\pi v_B} \f{p_f^3}{\Lambda^3}
                            \f{|\omega|}{v_f p_f} \right).
\end{equation}
This logarithmic behaviour 
leads to a logarithmic correction to the specific heat of the electron gas
\cite{HNP}. This result is not dependent on the choice
of gauge as can be seen by again computing
the contribution to the self-energy using the longitudinal part
of the gauge-propagator,
\begin{eqnarray}
 \Sigma_l(\vec{p},\omega_n) & = &
                              \f{g^2}{4 \pi^2} T \sum_m \int_0^\Lambda dq 
                              \f{\alpha}{q^2-\alpha B(q,\epsilon_m)}
                              \left[ i \omega_n-\f{p^2}{2m}+\mu+\mathrm{i} \epsilon_m 
        \right. \\
        &   &\left. +\f{m}{pq} 
         \left( \mathrm{i} \omega_n-\f{p^2}{2m}+\mu+\mathrm{i} \epsilon_m \right)^2
         \ln \left[ \f{-1+\f{m}{pq} 
             \left( \mathrm{i} \omega_n-\f{p^2}{2m}+\mu+\mathrm{i} \epsilon_m-\f{q^2}{2m}\right)}
            { 1+\f{m}{pq} 
             \left( \mathrm{i} \omega_n-\f{p^2}{2m}+\mu+\mathrm{i} \epsilon_m-\f{q^2}{2m}\right)}
             \right] \right]. \nonumber 
\end{eqnarray}
Expanding the logarithm we get
\begin{eqnarray}
 \Sigma_l(\vec{p},\omega_n) & = &
                              \f{g^2}{4 \pi^2} T \sum_m \int_0^\Lambda dq 
                              \f{\alpha}{q^2-\alpha B(q,\epsilon_m)}
                              \left[ 
         \left(\mathrm{i} \omega_n-\f{p^2}{2m}+\mu \right) 
         \nonumber \right. \\
        &   & \times
         \left(1+\mathrm{i} \epsilon_m \f{m}{pq} \mathrm{i} 
               \pi \mathrm{sgn}(\omega_n+\epsilon_m) \right)
         \nonumber \\
        &   &+\mathrm{i} \epsilon_m+(\mathrm{i}\epsilon_m)^2 \f{m}{pq} 
         \left( \mathrm{i} \pi \mathrm{sgn}(\omega_n+\epsilon_m)+\cdots \right)
           \nonumber \\
        &   &+\left. O\left( \mathrm{i} \omega_n-\f{p^2}{2m}+\mu \right)^2
                  \right].
\end{eqnarray}
The most singular term contributing to $Z_m$ is
\begin{equation}
  \Sigma^{Z_m}_l(\vec{p},\omega_n) = 
   \f{g^2}{4 \pi^2} \f{m}{p} T \sum_m 
    \int_0^\Lambda dq \f{\alpha q^{-1} (\mathrm{i}\epsilon_m)^2 \mathrm{i}\pi 
    \mathrm{sgn}(\omega_n+\epsilon_m)}{q^2-\alpha B(q,\epsilon_m)}.
\end{equation}
Taking the zero-temperature limit and using the expression (\ref{d3B}) for $B(q,\epsilon)$, we get in the limit $\omega \rightarrow 0$
\begin{equation}
   \Sigma^{Z_m}_l(p_f,\omega) = - \mathrm{i} \omega 
                                  \f{\pi}{16} 
                                  \left( \f{\alpha g^2 v_f}{2\pi^2} \right)^{1/2}
                                  \f{|\omega |}{v_f p_f}.
\end{equation}
This vanishes faster than $\omega$ as $\omega \rightarrow 0$, so there are
no gauge-dependent corrections to the singularities of $Z_m$ in this case 
either.
\section{Conclusion}
There have been many attempts \cite{NW}-\cite{KHM} to go beyond the mean field
approximation for the $\nu=1/2$ Quantum Hall state. Some
of these attempts \cite{NW},\cite{CNS},\cite{Stern} are based on
the behaviour of the singe-particle Green function. 
One criticism put forward \cite{KFWL} against arguments
based on the single particle Green function is that conclusions
based on it might not be physical since the single-particle
Green function is not gauge-invariant. \\
\\
In this paper we have shown that the dominant singular behaviour
determining the pole-structure of the single-particle
Green function is not affected by the choice of gauge,
when the gauge choices are restricted to a particular family of
gauges. This is true both in $d=2$ and in $d=3$. 
\\
\\
 This implies in particular that the $\beta$-functions obtained
in \cite{NW},\cite{CNS} are independent of gauge, and that 
the pole-structure used to construct the (singular) Fermi liquid
description in \cite{Stern} is also gauge-invariant.
\\
\begin{ack}
I wish to thank Sudip Chakravarty for his continuous 
intellectual support during this work, and Chetan Nayak for helpful
comments on the manuscript. This work was supported by the
National Science Foundation, NSF-DMR-92-20416.
\end{ack}

\end{document}